# Magnetic Materials via High-Pressure Torsion of Powders

Lukas Weissitsch [a, *] and Franziska Staab [b, *], Karsten Durst [b], Andrea Bachmaier [a]

[a] *Erich Schmid Institute of Materials Science of the Austrian Academy of Sciences, 8700 Leoben, Austria*
[b] *Physical Metallurgy (PhM), Materials Science Department, Technical University of Darmstadt, Alarich-Weiss-Str. 2, 64287 Darmstadt, Germany*
[*] Authors contributed equally
E-Mail address: lukas.weissitsch@oeaw.ac.at; franziska.staab@tu-darmstadt.de

**Graphical Abstract**

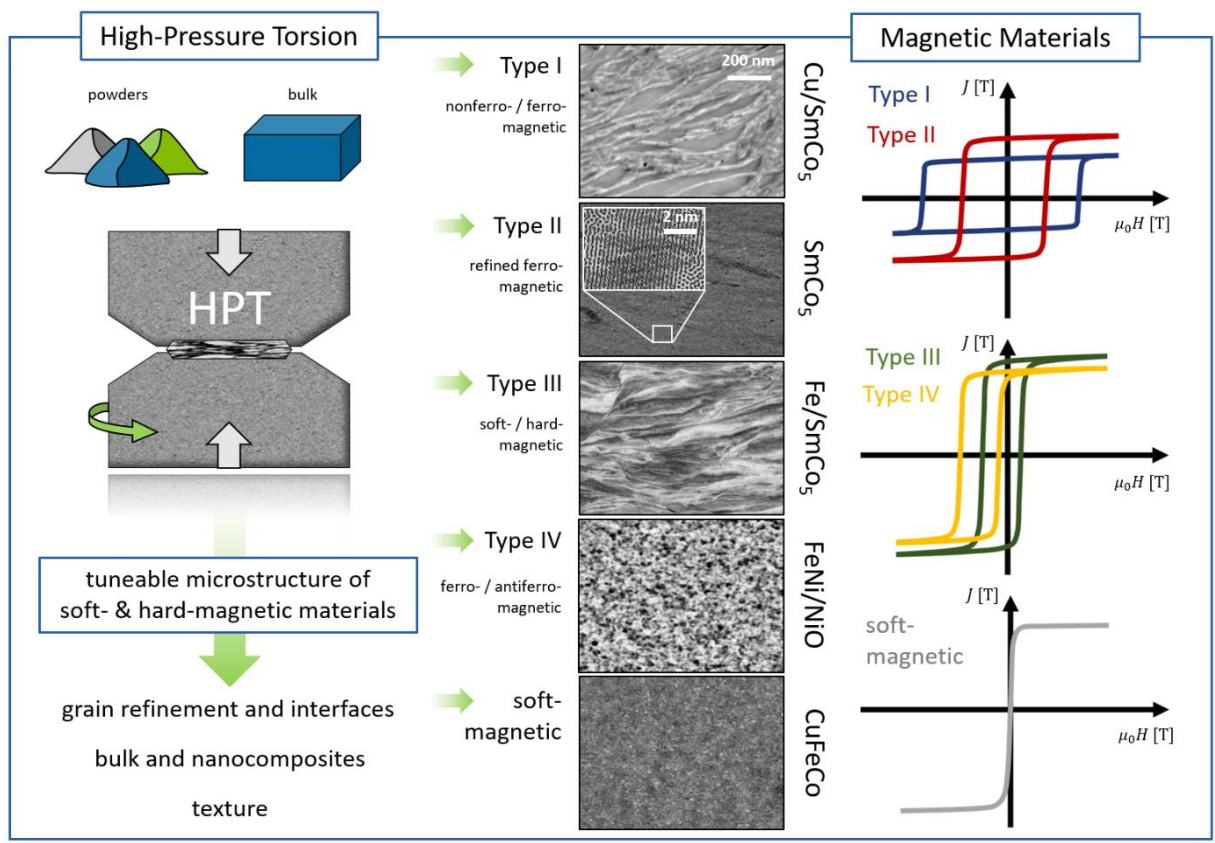

**Abstract**

Magnets are key materials for the electrification of mobility and also for the generation and transformation of electric energy. Research and development in recent decades lead to high performance magnets, which require a finely tuned microstructure to serve applications with ever increasing requirements. Besides optimizing already known materials and the search on novel material combinations, an increasing interest in unconventional processing techniques and the utilization of magnetic concepts is apparent. Severe plastic deformation (SPD), in particular by high-pressure torsion (HPT) is a versatile and suitable method to manufacture microstructures not attained so far, but entitling different magnetic coupling mechanisms fostering magnetic properties. In this work, we review recent achievements obtained by HPT on soft and hard magnetic materials, focusing on powder as starting materials. Furthermore, we give specific attention to the formation of magnetic composites and highlight the opportunities of powder starting materials for HPT to exploit magnetic interaction mechanism.

**Main text**

**Introduction**

Severe plastic deformation (SPD) is a well-known and effective method to produce ultrafine grained or nanocrystalline (nc) materials. One classic SPD method is high-pressure torsion (HPT), a novel process for the production and treatment of magnetic materials promising particularly noticeable magnetic properties. The grain refinement during HPT processing is accompanied with the formation of various kinds of lattice defects and unique microstructural features, but also to the formation of a shear texture.[1,2] HPT-deformation is widely used in processing many kinds of single-phase materials and also a variety of (nano)composite structures can be obtained.[3,4] Furthermore, the possibility for phase transformations[5–8] has been shown as well as the evolution of thermodynamically metastable phases such as supersaturated solid solutions[3,9] or amorphous structures, similar to rapid solidification processes.[10] This variety of HPT-induced microstructural evolutions is accompanied by a noteworthy manufacturing advantage. As initial materials for HPT processing, coarse-grained multi-phase alloys, powder blends of different elements or any other combination of solid starting materials can be used.

The HPT process applied to powder blends represents a tool for possible mixing, consolidation and subsequent alloying obtaining a solid solution or nanocomposite material by the high amount of applied shear strain. Thus, additional processing steps (e.g. sintering) to achieve a bulk material are not necessary. The torsional shear straining under high hydrostatic pressures by HPT can prevent the formation and propagation of cracks, so that also brittle materials, such as most of the hard magnetic phases, can be plastically deformed up to large strains. This is specifically of importance for the consolidation, refinement and plastic deformation of powder mixtures, where large hydrostatic stresses are required to form bulk samples from powders.[11] The application of powders enables a free selection of the elements or rather phases, resulting in no limitations by phase diagrams regarding metallurgical processing routes, due to the limitation of phase formations from a melt. The process hence represents a tool for mechanical alloying, phase transformations and supersaturation, often accompanied with pronounced grain refinement, while still resulting in bulk materials. The free selection of the starting powders is especially interesting, since thermodynamically metastable chemical compositions but also composite structures are achievable.

Further benefits of the obtained bulk samples from HPT are their complete density and large sample sizes. Typical HPT discs exhibit diameters of 10-50 millimeters, with the possibility of an upscaling, while simultaneously microstructural length scales are strongly reduced.[12] For example, the manufacturing of conventional permanent magnets on an industrial scale is based on powder metallurgical processing, requiring elaborate powder fabrication followed by sintering and further heat treatments. An excellent summary on processing routes for the most important permanent magnet systems can be found in reference.[13] A corresponding review[14] is addressed to the industrial importance of magnetic composites, so called bonded magnets, as well as to novel manufacturing techniques. The HPT process applied to powder blends resulting in bulk materials can go beyond the limitations of conventional processing routes, since the hard magnetic phase and the grain boundary phase can be chosen freely.

HPT-deformation is traditionally used to tune and study advanced mechanical properties. In recent years, a growing interest in functional characteristics of HPT materials is also apparent, since the microstructure strongly controls the magnetic properties of magnetic materials and can be further used to tune them. The performance of magnetic materials correlates with main advantages of HPT, i.e., grain refinement of single-phase materials, the possibility to form nanostructured composites or supersaturated solid solutions, while simultaneously texture formation through the deformation process is feasible.[15,16] Thus, an increasing interest of material scientists in the field of magnetism is recognized together with new possibilities for the development of magnetic materials. For instance, deformation induced metastable phases with large defect densities can be used as precursors for further processing, i.e., annealing treatments with enhanced dynamics or phase formations.

This article focuses on magnetic materials processed by HPT. It is organized as follows: Section 1 gives a short overview on intrinsic and extrinsic magnetic properties of hard and soft magnetic materials, considering the influence of microstructure and exchange interactions on the magnetic properties. Section 2 gives a short overview on bulk magnetic materials prepared by HPT. Section 3 is dedicated to HPT processing of magnetic powder blends and includes our own recent research on different magnetic materials. The last section contains an outlook on further research and summarizes the open questions in this field.

For further information and a more detailed discussion on dynamic interplay between microstructure evolution and magnetic properties, the interested reader is referred to some excellent review articles.[11,13,14,17–21]

1. **Magnetic mechanisms and microstructural tuning (microstructural tuning for magnetic hardening)**

The theoretical upper limit for achievable magnetic properties is governed by a combination of the intrinsic magnetic properties: magnetocrystalline anisotropy, exchange interactions, saturation magnetization ($M_s$) and Curie-Temperature ($T_C$). The exchange interactions are parameterized by the exchange stiffness ($A$), whereas the magnetocrystalline anisotropy is parameterized by the anisotropy constant ($K_1$). As schematically shown in Figure 1 the intrinsic curve is determined by the saturation magnetization $M_s$ and the anisotropy field $H_a$ giving the reversal field for a defect free single-domain Stoner-Wohlfarth particle by coherent rotation.[22,23] Using

$$H_a = \frac{2\,K_1}{\mu_0 M_s} \quad (1)$$

the anisotropy field ($H_a$) can be calculated.[22] If particles or grain sizes exceed the single domain particle size, domains are formed to minimize energy, in particular to reduce the stray field energy. However, the formation of a domain wall requires energy since the magnetic moments inside the domain wall do not align along their easy magnetization direction given by the magnetocrystalline anisotropy. Important characteristic length scales are the (Bloch) domain wall width $\delta_0 = \pi\sqrt{\frac{A}{K_1}}$ as well as the exchange length $l_{ex} = \sqrt{\frac{A}{K_1}}$ which are in the range of 2 to 5 nm for most ferromagnetic materials such as Fe, Ni, Co, SmCo$_5$, Nd$_2$Fe$_{14}$B.[22] Real materials do not obey the intrinsic hysteresis curve for a defect free single-domain Stoner-Wohlfarth particle, because the magnetization reversal processes are controlled by the microstructure. For instance, grain sizes, grain boundaries and any structural defects can act as nucleation sites for reversed domains but also hinder domain wall movement. Therefore, the measured hysteresis curve, called extrinsic curve, does not reach the upper limit, as depicted in Figure 1. Coercivity ($H_c$), $M_s$, remanent magnetization ($M_r$ or $B_r$) and rectangularity of the second quadrant (often characterized by the maximum energy product ($BH$)$_{max}$), which are technically important parameters of permanent magnetic materials, are hence strongly affected by the microstructure.

The discrepancy between the anisotropy field $H_a$ and the coercivity $H_c$, originating from defects and imperfections, is called Brown´s Paradox and is given by $H_c \leq H_a$.[22] The discrepancy and thus the real structural details of a magnet is usually described by the "Kronmüller factor" $α_K$, where $H_c$ equals $α_K H_a$.[24] In fact, it is difficult to obtain a $α_K$ larger than 0.3.[25] Therefore, besides good intrinsic properties, given by the comprised phases, a proper modification of the microstructure is required to tune the shape of the measured hysteresis loops and hence the magnetic properties. Microstructural features include the degree of texture (alignment of crystallites or grains), as well as the size distribution of the phase and precipitates. Also, other defects, like porosity or dislocations can influence the magnetization behavior.[8]

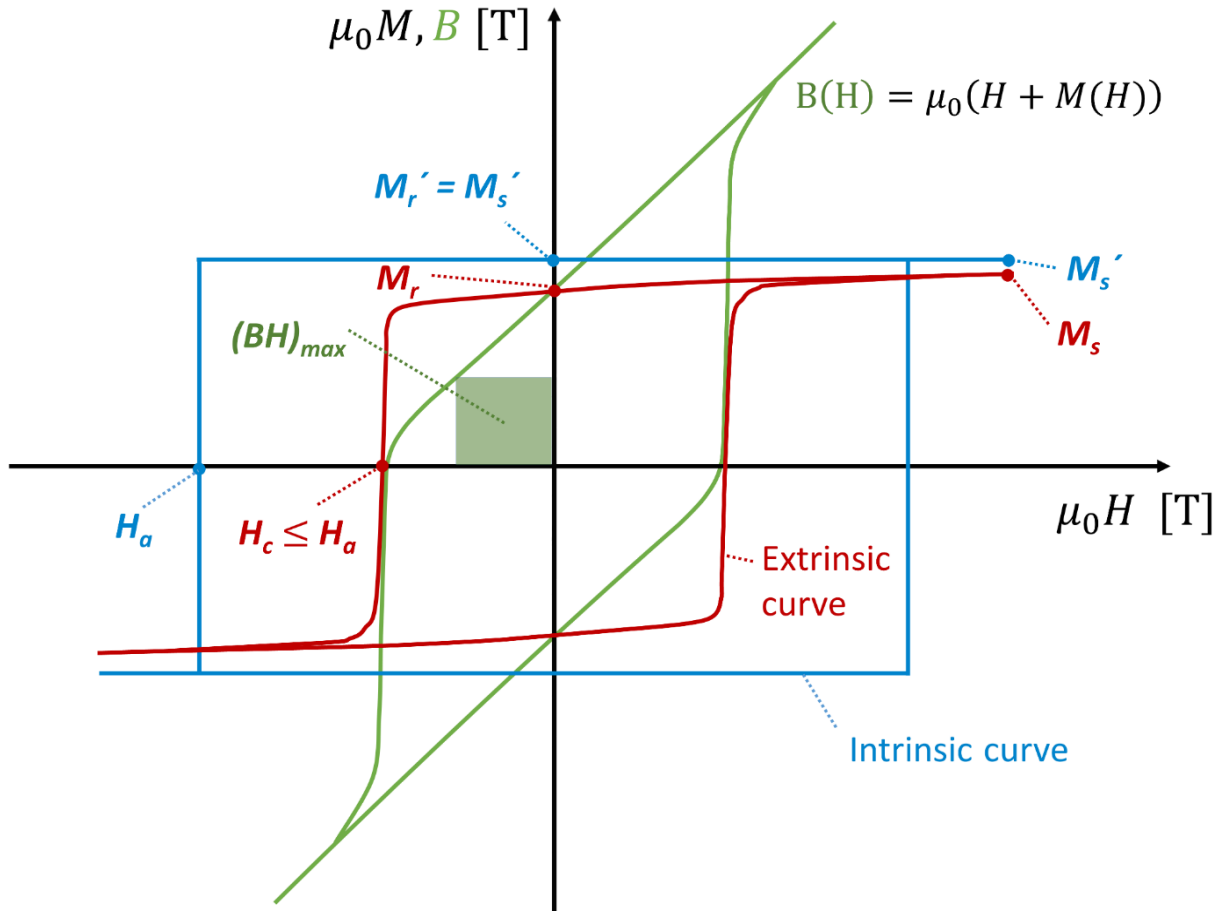

Figure 1: Schematic drawing of magnetic hysteresis: Magnetization µ₀M or magnetic flux density B as a function of applied field µ₀H. The Intrinsic properties of a material outline the theoretical upper limitation for magnetic materials (blue curve). Microstructural features decrease the upper limit of magnetic parameters (red curve) known as the Brown´s Paradox. Experimentally measured signals are frequently obtained in cgs-units, requiring thus additional measurements of the sample (e.g. density), to transform the units into SI-units. An extensive summary to transform magnetic units is published by the NIST.[26]

Depending on the microstructure two different coercivity mechanisms are distinguished. In nearly defect free permanent magnets nucleation of reversed domains controls the coercivity (nucleation-type magnets), whereas pinning of domain wall movement controls the coercivity in magnets with high defect densities (pinning-type magnets).[27] Pinning of domain walls can occur at various defects, which are often regions with a low anisotropy (small $K_1$) due to chemical disorder, such as precipitates, dislocations and grain boundaries. It is one major reason for the increase of $H_c$, often called as magnetic hardening, and most effective if the size of the pinning centers is comparable to the domain wall width $δ_0$.[27] Pinning of domain walls at dislocations is especially interesting for soft magnetic materials, due to their broad domain walls and the resulting increase of pinning strength.[28,29] Thereby, $H_c$ is given as $H_c \sim \sqrt{\rho_{VS}}$, with $\rho_{VS}$ the density of dislocations.[30] The reason for impeding domain wall motion is based on magnetoelastic interactions between the elastic stress field of dislocations and magnetic domain walls as well as on stress-induced anisotropy contributions.[30–32] Besides the coercivity also the initial susceptibility and the magnetization behavior are affected by dislocations in soft magnetic materials. Hard magnetic materials exhibit narrow domain walls leading to small pinning strength.[28,29] However, for the hard magnetic SmCo$_5$, Fidler et al.[33] reported a strong interaction between prismatic dislocations loops and domain walls.

Another microstructural feature which has a very crucial impact on the magnetic properties, especially the $H_c$, is the grain size. A more detailed description on the influence can be given when distinguishing soft and hard magnetic materials. This categorization is typically done by coercivity and the values for the two classes are given in Figure 2a).[21,34]

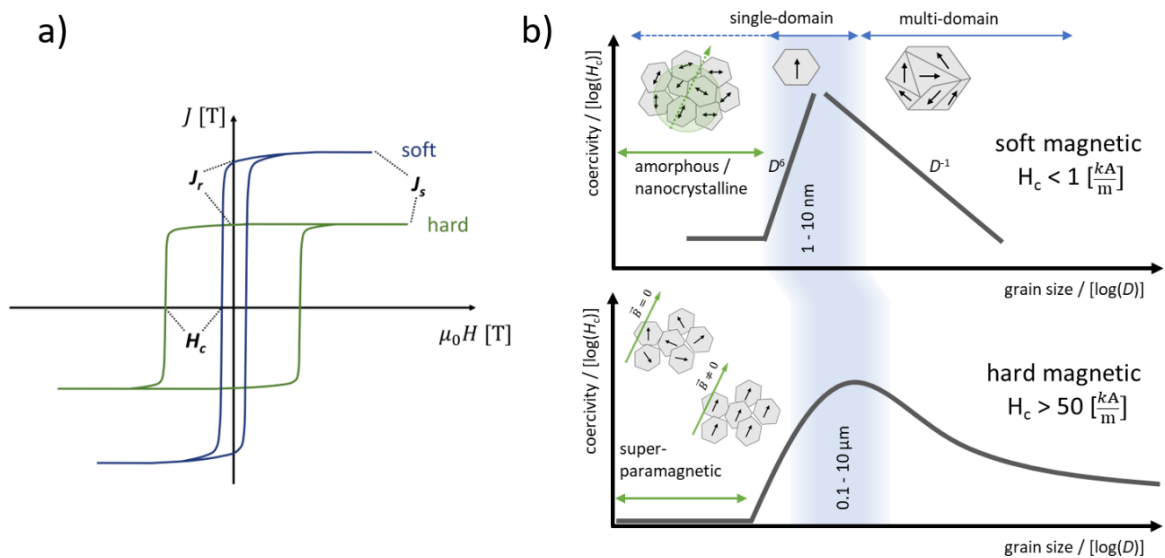

Figure 2: **a)** Magnetic hysteresis curve of soft and hard magnetic materials with marked characteristic values. **b)** Schematic representation of coercivity ($H_c$) as a function of the particle diameter (D) for soft (upper graph) and hard magnetic (lower graph) materials. Please consider the different ordinate axis scaling and the logarithmic application.  Adapted from [21,34–38]

The most prominent impact of grain sizes on coercivity is demonstrated by Herzer for soft magnetic materials (upper graph in Figure 2b). Therein, the $H_c$ scales inversely with the grain size until it approaches a maximum in the range of the single domain sizes. The single domain size is reached, when the energy required for the formation of a domain wall within the grain, exceeds the particle's stray field energy, this region is herein roughly indicated by the shading in Figure 2b). For even smaller grain sizes the coercivity strongly decreases by $D^6$. Essentially, when crystallites are further decreased in size and reach an amorphous state, soft magnetic properties are not necessarily enhanced and $H_c$ is not vanishing. Instead the so-called random anisotropy model explains a remaining coercivity which is based on the exchange length $L_{ex}$, depicted as a green circle within Figure 2b). If $L_{ex} \gg d$ is fulfilled, within this length scale, a statistical averaging of the anisotropic moments leads to the formation of a non-vanishing $H_c$. Some commercial soft magnetic materials are also nanocomposites, in which a crystalline phase is embedded in an amorphous matrix and the model of Herzer has been successfully applied to describe their coercivity behavior. Most importantly, if there is a long-range order of the amorphous structure and at the same time a uniaxial anisotropy is present, the dependency on coercivity is changed from $D^6$ to $D^3$.[39]

For hard magnetic materials, the drawn characteristics in the lower part of Figure 2b) shows a tendency of a maximum in coercivity close to the single domain size, without a defined power law dependency on the grain size. Most significantly, due to their higher magnetocrystalline anisotropy, the single domain size is increased. The coercivity dependence on grain size does not allow a strict proportionality as depicted in the upper part of Figure 2b) since all possible microstructural features, but also the interplay of intrinsic and extrinsic properties or chemical variations contribute separately and in a different extent. A widely known model is the one by Ramesh et al. which expresses $H_c$ as a function of $\ln(D)$ for grain sizes larger than 3 µm. The model suggests a higher nucleation probability of reversing domains for larger grain sizes.[40,41] A conclusive model is still lacking, thus several explanations are present.[42,43] Nevertheless, several observations, i.e., experimental data from Lee et al.[44] suggest a grain sizes dependency on $H_c$. A further decrease in grain size well below the single domain size, at which a maximal $H_c$ is present, is again usually accompanied by a decreasing $H_c$, reaching a superparamagnetic state. Therein, magnetic moments directly align with the applied field, as they are completely separated and do not show any exchange interaction. As a result, the obtained hysteresis loop shows a vanishing $H_c$ and consequently no $B_r$, but can reach a magnetic saturation at sufficiently high fields. Counterintuitively, for soft magnetic materials the random anisotropy model increases $H_c$ compared to hard magnetic materials, in which the superparamagnetic state is present. However, the superparamagnetic state can also be reached in soft magnetic materials decreasing the coercivity e.g. when applying temperature.

For permanent magnets the maximum stored energy $(BH)_{max}$, which is defined as the maximum product of $B$ and $H$ in the second quadrant of the $B$-$H$-curve, is of critical interest and should be as high as possible. Therefore, a high $H_c$ and a high degree of texture is necessary in order to reach maximum values for $B_r$ and to increase $(BH)_{max}$. Depending on the microstructure two different coercivity mechanisms are distinguished as described above. For nucleation-type magnets, the coercivity reaches its maximum value at grain sizes at which only a single domain state is present. If a particle consists of several domains and an external magnetic field is applied, a domain having a favorable orientation of magnetization will grow at the expense of a poorly aligned domain. If only one domain is present in a particle, the alignment along the external field will occur via rotation of the magnetization direction, which usually costs more energy compared to growing of favorable aligned domains due to the magnetocrystalline anisotropy. This is the reason for the maximum coercivity at single domain grain sizes.

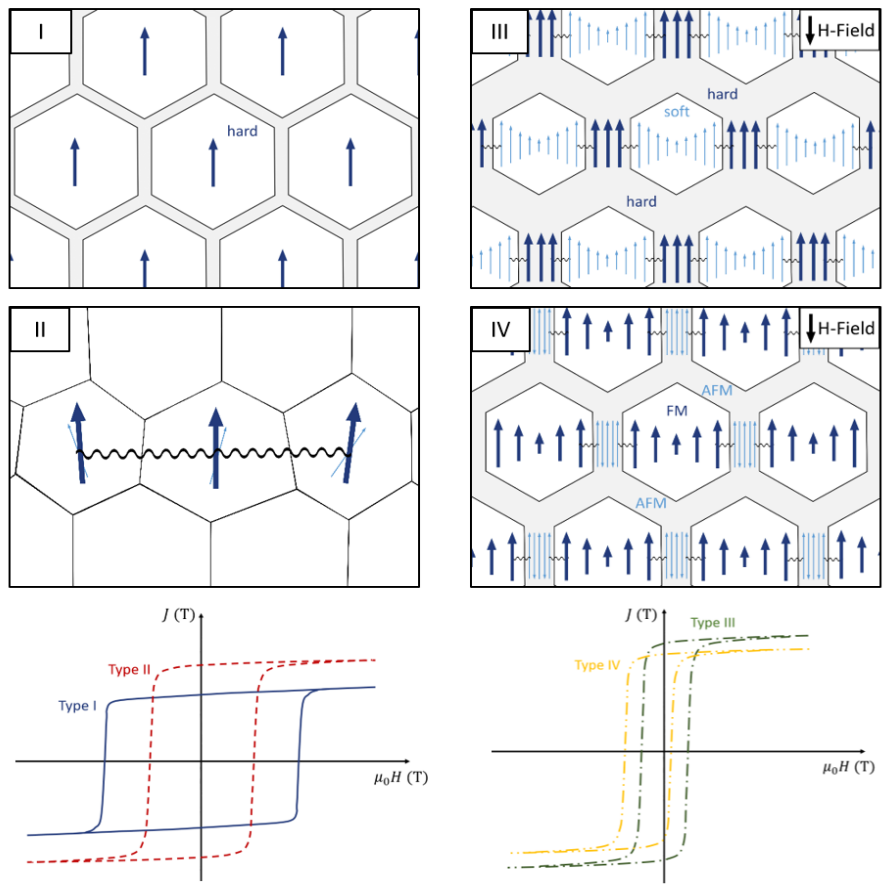

*Figure 3: Basic schematic representations of microstructures of nucleation-type permanent magnets with the corresponding characteristic magnetic hysteresis curves. Type I: decoupled isolated single domain particles, Type II: exchange interaction of particles, Type III: exchange coupled soft and hard magnetic phases, Type IV: exchange bias due to coupling of ferro- and anti-ferro magnetic interfaces. Adapted from [34,45]*

There are different basic schematic types of microstructures for nucleation-type magnets leading to different magnetic properties as depicted in Figure 3 together with the shape of their corresponding idealized hysteresis loops.[34] The type I microstructure leads to the highest coercivity and consists of aligned single domain ferromagnetic particles, separated by a thin non-magnetic layer. This layer magnetically decouples the individual ferromagnetic particles, which hence behave as single permanent magnets. In order to magnetically decouple the individual ferromagnetic particles, the non-magnetic grain boundary phase needs to be thicker than the exchange length to suppress exchange interactions between the magnetic moments of neighboring particles.

If there is no second phase and the microstructure only consists of small ferromagnetic particles, type II is obtained. The ferromagnetic particles are exchange coupled within the surrounding grains, leading to a vectoral alignment of the magnetic moments close to the grain boundaries and thereby to an increasing remanence. If, in addition to the hard magnetic phase a second soft magnetic phase is added, type III microstructure is achieved. Exchange coupling between the different phases, allow a further increase of the remanence due to the high saturation magnetization of the soft magnetic material. Because the hard magnetic phase stabilizes the soft magnetic one against magnetization reversal, this type of microstructure is referred to exchange coupled spring magnet.[46]

Using a composite consisting of a ferromagnetic and an antiferromagnetic phase and cooling the material in the presence of a static magnetic field from a temperature above the Néel-temperature to one below it (while the temperature treatment remains below the Curie-temperature of the ferromagnetic phase) leads to an exchange bias, as shown for the microstructure type IV. The exchange interactions at the interfaces between the antiferromagnetic and the ferromagnetic phase inhibit the magnetic spins of the ferromagnetic phase to rotate, leading to a shift of the hysteresis curve along the field axis of field cooling.[47]

Conventional microstructures often exhibit more than two phases or a more complex grain structure with changing chemical compositions ("core-shell" structures) could be present resulting in different mixed types. The complexity of microstructures and hence, magnetic interactions is expected to further increase in future. However, Figure 3 represents the basic simplified schematic interactions, which can be adapted accordingly.

Besides the different phases and the grain size, also the texture is very crucial, especially for hard magnetic materials. To obtain high values for $(BH)_{max}$, anisotropic magnets are required, where all grains should be aligned along their easy magnetization direction and the volume fraction of the hard magnetic material should be as high as possible.[48] The conventional production route for rare earth-transition metal-based permanent magnets such as NdFeB and SmCo is based on the powder-metallurgical sintering route.[13] Due to the particularity of the $Nd_2Fe_{14}B$ phase, having an anisotropic grain growth perpendicular to the c-axis, di-upsetting leads to highly anisotropic Nd-Fe-B magnets. However, this feature is not present in other compounds with excellent intrinsic magnetic properties. For these, bonding of nanocrystalline powders by a polymer, which are usually aligned in an external field, is another method of producing bulk permanent magnets.[14] However, by using polymers the amount of magnetic material and consequently $B_r$ and $(BH)_{max}$ are reduced.

Therefore, to generate permanent magnets, there is a challenge to find new processing techniques in order to achieve a microstructure with highly textured, well-aligned single domain particles of different compounds exhibiting excellent intrinsic magnetic properties. HPT-deformation can be used for the production and tuning of magnetic materials by controlling various microstructural features. It is a promising process which can go beyond the conventional processing techniques.

## 2. High-pressure torsion of bulk magnetic materials

Decreasing the grain size to the nc regime can drastically improve soft magnetic properties (drastic reduction in coercivity and lower losses) due to exchange coupling.[35,49] Although HPT on magnetic materials such as Fe, Co and Ni were already reported in 1935 by Bridgman, the first works considering the magnetic properties of these HPT processed materials were published in the 1990ties.[50,51] However, in HPT-deformed bulk single-phase materials (e.g. Co, Ni, Fe, Fe-Si alloys), soft magnetic properties could not be fully achieved and often an increase in coercivity is reported.[50–58] One of the reasons is a too large saturation grain size in single phase materials, which means that the power law dependence of the coercivity on the grain size, as shown in Figure 2, was never fulfilled. Another important point that hinders the achievement of truly soft magnetic properties is the high density of defects, which are introduced during HPT-deformation.

HPT processing decreases microstructural feature sizes, which makes it possible to additionally increase coercivity in hard magnetic materials (type II hard magnetic material, c.f. Figure 2b). Several groups report a coercivity enhancement in different hard magnetic materials due to grain refinement after HPT processing, compared to the initial state.[59–63] In addition to grain refinement in some HPT-deformed hard magnetic materials, a (partial) amorphization has been observed.[59,64–67] In[65], for example an amorphous structure was induced in Nd(Pr)–Fe–B alloys after HPT to large strains. Subsequent annealing resulted in recrystallization, the formation of stable magnetic and non-magnetic phases. In consequence, significant improvements in hard magnetic properties were reported. Another approach is to deform already amorphous magnetic materials by HPT.[68–73] In[68], for example, the effect of HPT-deformation and subsequent annealing on structure and magnetic properties of overquenched melt-spun $Nd_9Fe_{85}B_6$ alloy was investigated. During HPT-deformation, α-Fe nanograins are formed which act as nucleation sites for $Nd_2Fe_{14}B$. Thus, uniform and nc α-Fe and $Nd_2Fe_{14}B$ grains formed during subsequent annealing improving coercivity and remanence (type III hard magnetic material, c.f. Figure 2b). A similar influence on microstructural development and improved magnetic properties were observed in.[69,70,72–78] HPT-deformation can further lead to phase decomposition during deformation in hard magnetic alloys.[79] For a $Sm_2Fe_{17}N_x$ alloy[79], phase decomposition in α-Fe and SmN was observed, which lead to an exchange coupled type III hard magnetic material with increased coercivity, remanence and saturation magnetization. Recently, the microstructural changes of the $Sm_2Fe_{17}N_3$ compound upon HPT-deformation were further studied by Hosokawa et al.[80] HPT-deformation can also enhance the kinetics of phase formation and/or transformations. For example, the synthetization of L1$_0$-FeNi tetrataenite during HPT-deformation and subsequent annealing has been studied in.[81–87] Up to now, the only successful L1$_0$-FeNi phase formation using HPT-deformation has been reported in.[85–87] Additionally, the various defects introduced by HPT can act as pinning sites in hard magnetic materials, hindering domain wall movement and change the magnetization reversal process from nucleation to a pinning dominated one. For example, this effect has been observed in MnAl based materials, which is prone to twin formation during HPT-deformation.[62,63,68,88–92] However, the influence of extensive twinning on coercivity is discussed controversially in literature.[91]

### 3. HPT-deformation of magnetic powders

HPT-deformation of powders is as old as the HPT process itself, as already Bridgman applied it to powders in 1935.[93] Even though he used the term "self-welding" he reported consolidation in materials such as Y, Er[93] and borox glass.[94] Nowadays, besides the consolidation of metallic powders[95–97] also composites[98,99] have been generated by HPT applied to powder blends. As starting powders commercially available powders of single elements or alloys can be used as well as powders produced by ball milling or finely crushed ribbons.

As already discussed in the introduction, the free selection of the starting powders is especially interesting, during the HPT process using powders as starting materials, since thermodynamically metastable chemical compositions often interesting for soft magnetic materials, but also composite structures are achievable. This is of great benefit, in particular for hard magnetic materials, where the hard magnetic phase and the grain boundary phase can be chosen freely. The microstructures shown schematically in Figure 3 can be generated by the choice of the initial powders, the consolidation and deformation process (see Figure 4).

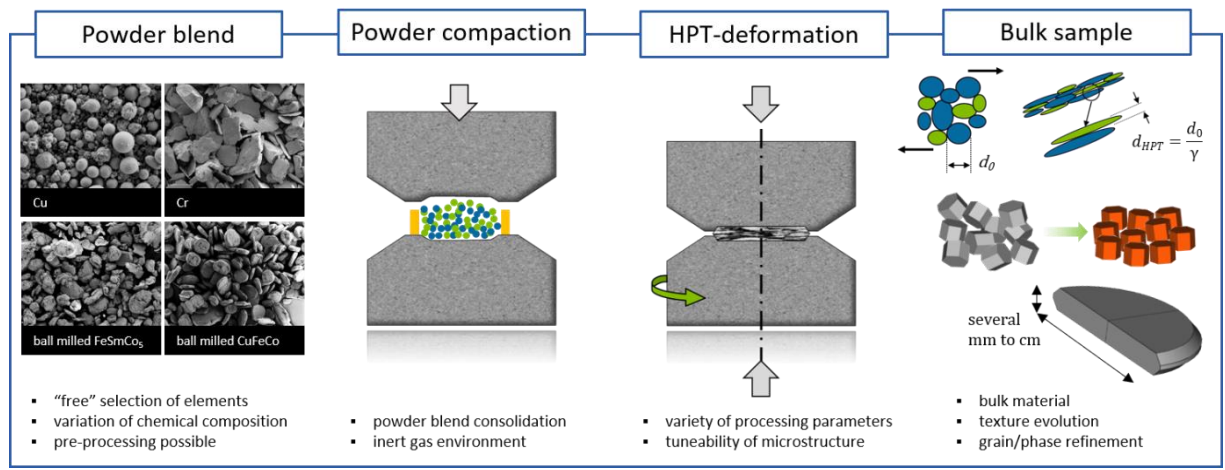

*Figure 4: Schematic representation of powder-based HPT processes. Usually any kind of powders or powder mixtures are consolidated by a compaction step. The obtained pellet is deformed by HPT resulting in bulk materials with enhanced microstructural features.*

A general HPT processing route of powders is schematically shown in Figure 4. In a first step the powder or powder blend is chosen, which can be pretreated in any kind. The powder gets poured into the lower anvil by the aid of a Cu-ring, which is stuck to the lower anvil as depicted in Figure 4. The mixture is pre-compressed, already often accompanied by a slight rotation and/or oscillation by a few degrees. This leads to cold compaction and the formation of a stable pellet. Depending on the materials but also on their reactivity due to an increased surface to volume ratio, oxidation sensitivity has to be considered. This can be impeded by using inert gas atmospheres for storage, handling and also compaction, if special equipment is used.[100] The actual HPT-deformation is done in a second step. By variation of different process parameters such as pressure, strain, strain rate and temperature, different microstructures and hence different magnetic properties can be adjusted. Compared to the deformation of bulk materials the HPT-deformation of powders usually leads to finer microstructures and a higher hardness. This behavior can be beneficial for mechanical or magnetic properties and is related to the higher amount of precipitates and oxide particles of powder materials.[3,97,101,102]

The variety of possible chemical compositions, desired microstructures and material phase distributions is accompanied with challenges within the HPT-processing. Intrinsic material parameters, their compatibility of plasticity as well as the phase composition often lead to an inhomogeneous morphology or a general brittle sample behavior.[3,103] Localization of the deformation is sometimes difficult to avoid and a microstructural saturation through the whole cross section of the HPT-sample could be unattainable due to the radial strain dependency and an enormous hardness increase at the HPT-disc edge.[12] Although, the high hydrostatic pressures during HPT, which is further increased by the cavity of the anvils - the so called constrained HPT setup - helps to circumvent these processing limitations. Therefore, a comparability of the applied shear strain between different samples is a challenging task to obtain. The additional variety of possible processing parameters include a change of the strain rate or the deformation at low or elevated temperatures, thus further investigations and improvements of processes and its parameters are needed.

### 3.1. HPT-deformation of powders to obtain soft magnetic materials

It is well known that materials with supersaturated phases, synthesized by HPT from alloy compositions immiscible at the thermodynamic equilibrium, yield smaller grain sizes compared to single phase materials.[2] This is also true for powder mixtures consisting of ferromagnetic and diamagnetic elements (e.g. Co-Cu, Fe-Cu, Fe-Ag).[104,105] In these HPT-deformed samples, an initial multiphase structure vanishes as a saturated state (steady state microstructure) is reached. This usually is obtained for shear strains above 1000, which equals to about 50 to 150 HPT rotations for a steady state regime, depending on the sample dimensions and the considered radius.[12] Subsequent annealing induces demixing and grain growth. Thus, annealing treatments can also be used as a further tool to tune magnetic properties.[106]

For example, fcc single phase nanocrystalline microstructures were achieved in Co-Cu alloys (similar to type I microstructure of permanent magnets), where the grain sizes decreased from 100 nm to 77 nm with increasing Co contents.[104] APT and TEM investigations clearly show that a homogeneous Cu and Co concentration is obtained even in nm regions.[106,107] Paramagnetic features for Co contents less than 50 wt. % (Figure 5a) and a decrease of the magnetic moment per Co-atom with Cu-content, which might originate from Ruderman–Kittel–Kasuya–Yosida-interaction, were observed with superconducting quantum interference device (SQUID) hysteresis measurements.[108] The microstructure remained single phase up to annealing temperatures of 400°C, but higher annealing temperatures induced phase separation into pure Cu and Co. The measured coercivity increased with increasing Cu-content and annealing temperature (Figure 5b and c).[106,107] Transmission-Kikuchi diffraction analysis revealed that the grain size is still above the exchange length of bulk fcc-Co.[109] Thus, the coercivity only qualitatively coincides with the random anisotropy model (Figure 5d). Using the micromagnetic constants of pure Co to explain the measured data fails.[110] Thus, micromagnetic constants of supersaturated solid solutions need to be calculated and are the focus of current research.

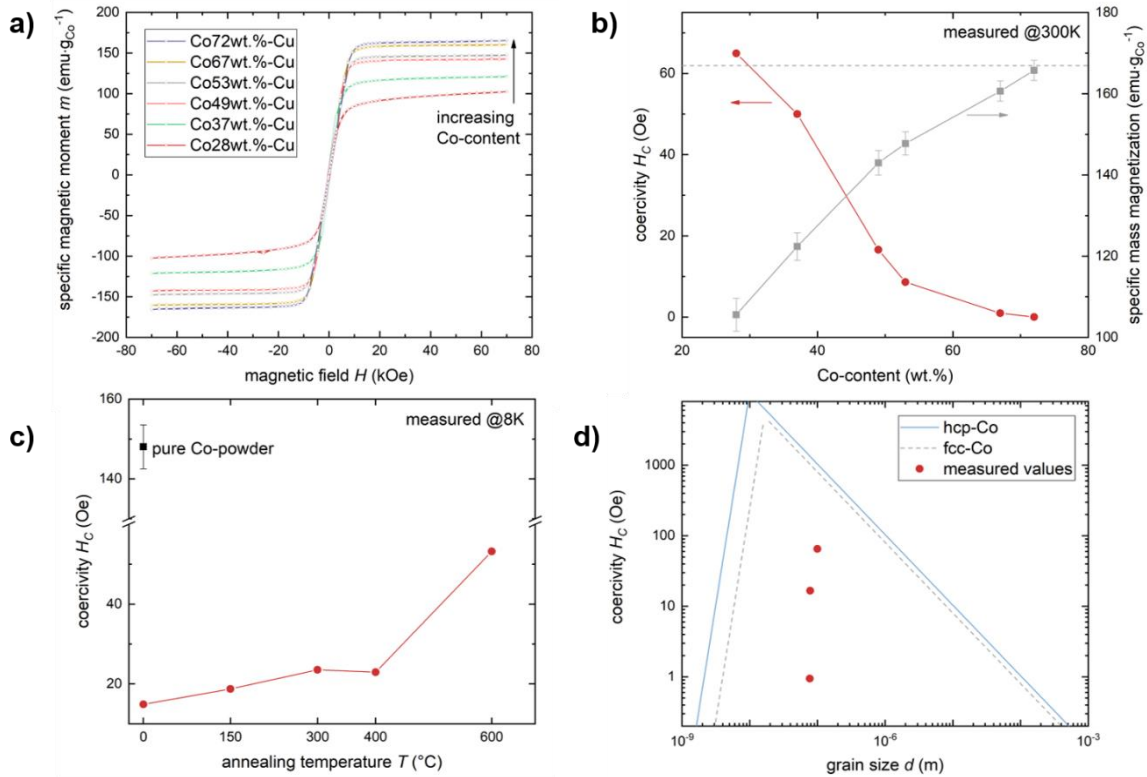

*Figure 5: **a)** Magnetization versus applied field for Co-Cu samples with different compositions measured at room temperature. **b)** Coercivity and magnetization versus Co-content measured at room temperature. **c)** Coercivity as function of annealing temperature for Co72wt.%-Cu. **d)** Herzer plot using micromagnetic constants of bulk fcc and hcp Co (solid lines) and experimentally obtained values from HPT-deformed Co-Cu samples. Adapted from reference.[106]*

Another ferromagnetic-diamagnetic system, which has been investigated, is Fe-Cu. It is very attractive because Fe has a high magnetic moment and low costs.[22] Using different sample preparation methods, it has already been shown that soft magnetic properties with a tunable magnetic hysteresis can be obtained in supersaturated solid solutions.[110–114] The Fe-Cu alloys synthesized by HPT-deformation consisted mainly of fcc-Cu phase with tiny, residual Fe particles.[107] Detailed APT investigations showed that Fe-clusters are present in the as-deformed microstructures. At the same time, Fe is also dissolved in the fcc-Cu grains. The ratio of clustered to dissolved Fe varied as a function of composition, which further changed the magnetic properties as revealed by SQUID magnetometry.[107] In Figure 6a), the DC-magnetization versus applied magnetic field for samples with increasing Fe content measured at 8 K and room temperature are shown. No saturation of the hysteresis is observed even at the highest applied field. The formation of a supersaturated solid solution of Fe in Cu is confirmed by the measured non-zero high-field susceptibility, which is due to a suppression of the ferromagnetic long-range interaction. Coercivities measured in the as-deformed state and its evolution with annealing quantitatively complies with the random anisotropy model similar to the Cu-Co alloys. Normalized zero-field cooling (ZFC) and field-cooling (FC) curves of as-deformed samples as a function of Fe-concentration are plotted in Figure 6b). The Fe25 wt.%-Cu samples showed a typical thermally activated behavior (superparamagnetism due to the residual Fe particles). The Fe14 wt.%-Cu sample displayed splitting of both curves and also a maximum in the FC-curve (at about 50 K). A similar maximum, but no splitting, was observed for Fe7 wt.%-Cu sample. That behavior was correlated to a magnetic frustrated state (spin-glass behavior of diluted Fe), which proved that deformation-induced mixing of Fe and Cu by HPT can be achieved on an atomic scale.[115,116] In the Fe-Cu sample with medium Fe content (14 wt.%), the magnetic frustrated state was proven by testing for the dynamic scaling law for spin-glasses[35,53] and by magnetic-field shifts following the Almeida-Thouless line[57] using frequency dependent AC susceptibility measurements in the as-deformed state and after annealing. Both magnetic phases (frustrated and thermal activated) are still present after annealing at low temperatures (150°C). The coercivity decreases from 128 Oe in the as-deformed state to 55 Oe after annealing at 150°C, which indicates a reduction in residual stresses. During annealing at 250°C, the magnetic frustrated phase vanishes, but larger Fe particles in the microstructure increase the coercivity. After annealing at 500°C, a bulk ferromagnetic behavior was observed, which indicated the formation of multidomain particles confirmed by microstructural investigations.[106]

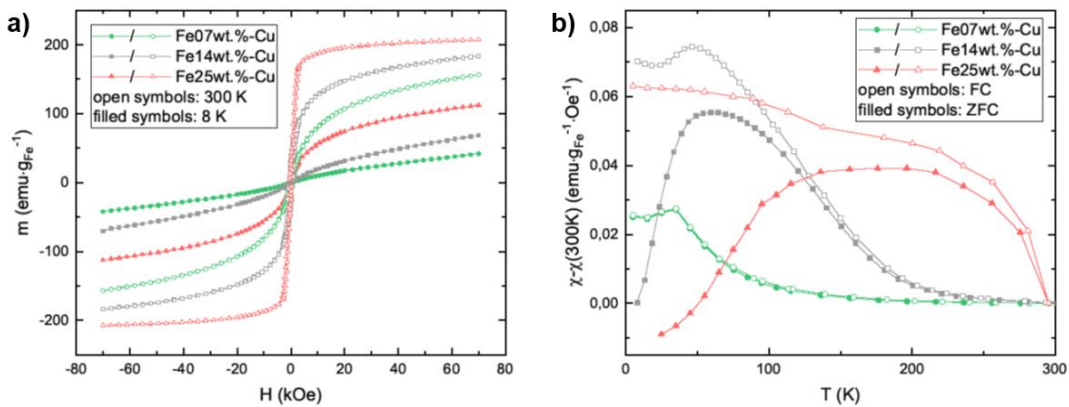

*Figure 6: **a)** Magnetization versus applied field for HPT-deformed Fe-Cu samples measured at room temperature. **b)** Normalized ZFC (open symbols) and FC curves (filled symbols) with an applied field H=50 Oe of as-deformed samples as a function of Fe-content. Adapted from reference.[107]*

The results show that in the HPT-deformed immiscible systems magnetic tunability is possible by changing the chemical composition or by subsequent annealing and semi-hard magnetic properties are reached. To push the limits towards enhanced soft magnetic properties, different Fe-Co ratios are mixed with Cu in Ref.[117] as the addition of Fe can lead to decreasing magnetocrystalline anisotropy and increasing saturation magnetization. Based on findings from binary compositions deformed by HPT-deformation, a region in the ternary phase diagram to synthesize homogeneous nanocrystalline supersaturated solid solutions has been identified (Figure 7). For samples containing a higher ferromagnetic content, a two-step HPT-deformation process at different temperatures was used. Depending on the ratio of Fe and Co, samples with a high ferromagnetic content either form a single-phase supersaturated solid solution or a microstructure which consist of bcc- and fcc-phases. Soft magnetic properties were confirmed for large Co-to-Fe-ratios. For example, the coercivity of a Co-rich $Cu_{17}Fe_{11}Co_{72}$ is about 2 Oe at room temperature.[118] This low coercivity might originate from a decreased magnetocrystalline anisotropy due to the addition of Fe and might be further lowered by the formation of the supersaturated solid solution.

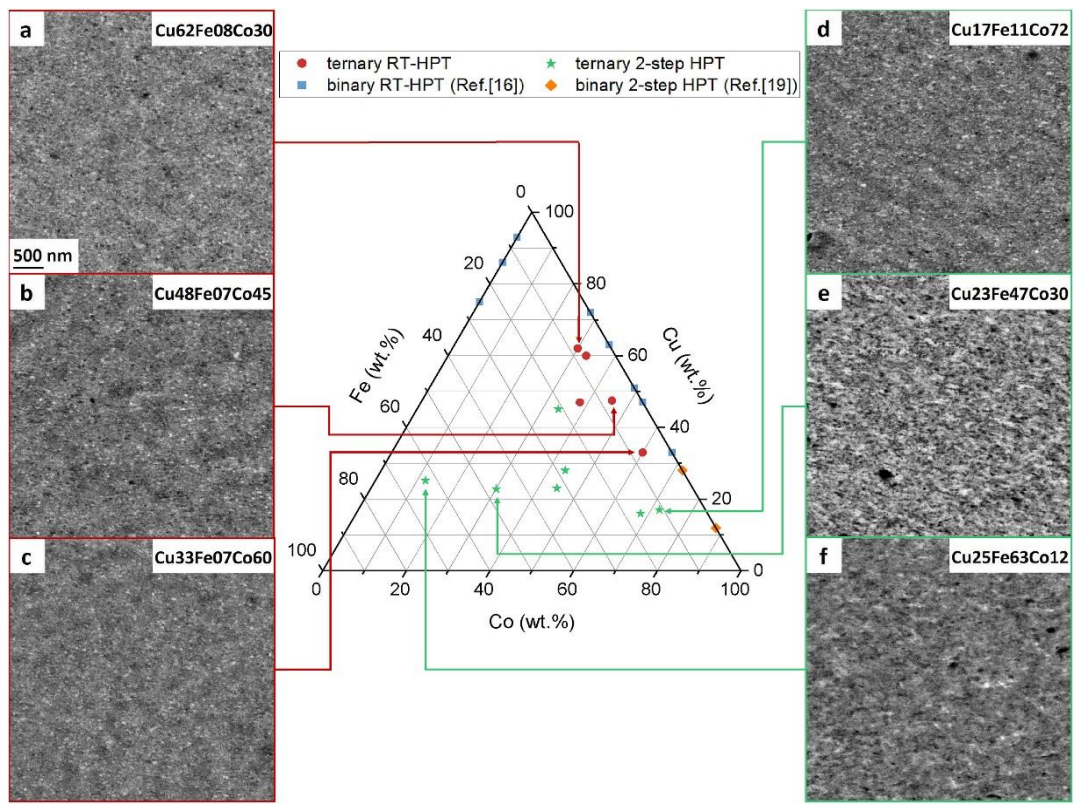

*Figure 7: The ternary phase diagram with compositions investigated in this study. Images on the left **a)–c)** show SEM micrographs of samples, which are processed by HPT-deformation at room temperature. Images on the right **d)–f)** show SEM micrographs of samples processed by the two-step HPT-deformation process at different temperatures. The scale bar in **a)** applies to all micrographs. Reprinted from reference.[117]*

Considering application, a further important magnetic property is magnetoresistivity. A giant decrease of resistivity with increasing applied magnetic field, the giant magneto-resistance (GMR) effect, was first observed for very thin layers of alternating ferromagnetic Fe and antiferromagnetic Cr.[119,120] The GMR phenomenon is not limited to layered systems and can be also found for materials which consist of nanometer sized ferromagnetic particles dispersed in non-magnetic metallic matrixes ('granular'

GMR).[121,122] Binary and ternary powder mixtures consisting of different compositions of elemental Cu, Co, Ni and Fe were thus processed by HPT-deformation and examined as possible candidates for the granular GMR effect.[123–125] In these studies, it could be shown that it is possible to produce HPT-deformed samples showing GMR behavior at room temperature. The GMR-ratio is small compared to multilayer systems already in application. It is, however, in a similar range to literature values of other materials processed by different methods showing granular GMR.[121,122] For example, a maximum GMR of ~3.5% was found for a Cu-Co sample with approximately equiatomic fraction of Co and Cu. For the investigated ternary systems, the highest drop resistivity at room temperature was found in $Cu_{62}Fe_{19}Ni_{19}$ after annealing (2.45% at 1790 kA/m). However, no saturation was measured in any composition and higher magnetic fields or lower measurement temperature are expected to lead to a stronger GMR effect.

### 3.2. HPT-deformation of powders to obtain hard magnetic materials

In the literature, HPT processing of powders has been successfully used for the production of NdFeB based materials.[126,127] An amorphous precursor was obtained from crushed melt spun ribbons and was consolidated applying a hydrostatic stress in the HPT device. During subsequent HPT-deformation, decomposition of $Nd_2Fe_{14}B$ to an amorphous matrix phase with finely dispersed nanocrystalline a-Fe-grains occurs.[126] In reference[127] a beneficial (00l) phase texture of $Nd_2Fe_{14}B$ upon HPT-deformation further induced anisotropic magnetic properties and resulted in a maximum energy product of 22.8 MGOe.

Based on $SmCo_5$ hard magnetic powder, it was demonstrated how the different magnetic exchange interactions and magnetic hardening mechanisms (c.f. Figure 3) can be successfully implemented by HPT-deformation of powders. Therefore, powder blends of $SmCo_5$ and magnetically different binder phases, i.e., Fe, Ni, Co, Cu, Zn, Sn and Cr, were applied to HPT-deformation. The microstructures in and corresponding hysteresis loops are summarized in Figure 8a) and b), respectivelyFigure 9. The different types of microstructure introduced in Figure 3 could be well reproduced. A type I magnetic system was obtained using diamagnetic Cu as a binder phase.[16] In preliminary experiments, type II magnets deforming SmCo-based intermetallic powders ($SmCo_5$, $Sm_2Co_7$, $Sm_2Co_{17}$) by HPT were fabricated as well. An exchange coupled type III system was prepared using Fe[128] and a type IV ferro-antiferromagnetic composite is represented by $SmCo_5$+Cr. Unfortunately, an exchange bias for $SmCo_5$ using Cr as binder phase (type IV system) could not be proven yet. Apart from exchange bias samples, the magnetic behavior follows the theoretical prediction. $M_s$ is highest for the Fe containing composition (exchange coupled spring magnet) and lowest for the diamagnetic and antiferromagnetic binder phase. $H_c$ is lowest for the type III sample, containing the lowest amount of SmCo$_5$, while $H_c$ is higher for the refined structure and further increases by adding Cr and Cu. The highest $H_c$ is obtained by the type I system, where Cu is magnetically isolating the refined SmCo$_5$ particles.

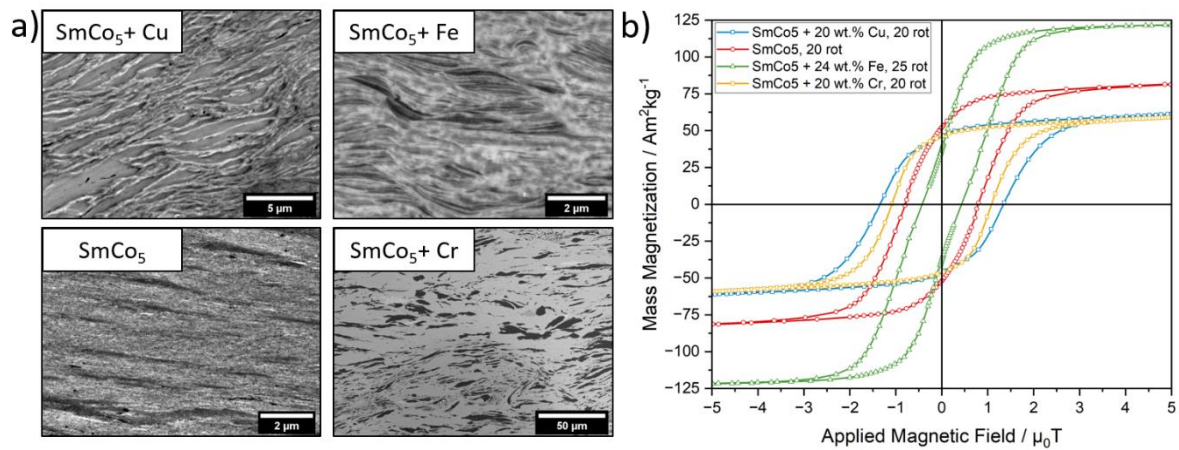

*Figure 8: **a)** BSE images of HPT-deformed samples at a radius of 2 mm. Different SmCo$_5$ based powder blends are used as starting material. **b)** Corresponding hysteresis measurements depicted between ± 5 T.*

The type I [16] and type III [15,128] SmCo$_5$ based material systems were investigated in more detail with a focus on the material processing of several chemical compositions and the influence of applied deformation on the microstructure. Staab et al.[16] deformed several samples to different extent, the composite structure was refined and an elongation of the hard magnetic phase particles, separated by Cu was reported. The best processability and simultaneously very good magnetic properties were found using 20 wt.% of Cu as binder phase. Increasing the amount of Cu did not lead to a deterioration of the coercivity but the diamagnetic Cu phase led to a dilution of the total magnetization of the sample. A similar microstructure evolution was apparent for Fe-SmCo$_5$ composites as presented by Weissitsch et al.[128] Figure 9a)-d) shows BSE images of a RT deformed 47 wt.% Fe - SmCo$_5$ composite at different radii of a single sample. A mainly compression dominated deformation is visible at radius 0 (Figure 9a). With increasing radius, the inhomogeneous structure with separated phases of Fe (dark contrast) and SmCo$_5$ (bright contrast) remains. However, the distance between the phases decreases and a lamellar morphology, well below 1 µm is found for radius 3 in Figure 9d)-e). Refined heterophase structures are observed for both material systems.[15,16,128] However, if ball milled powders were used, the resulting phase morphology changes drastically depending on the used chemical composition.[15]

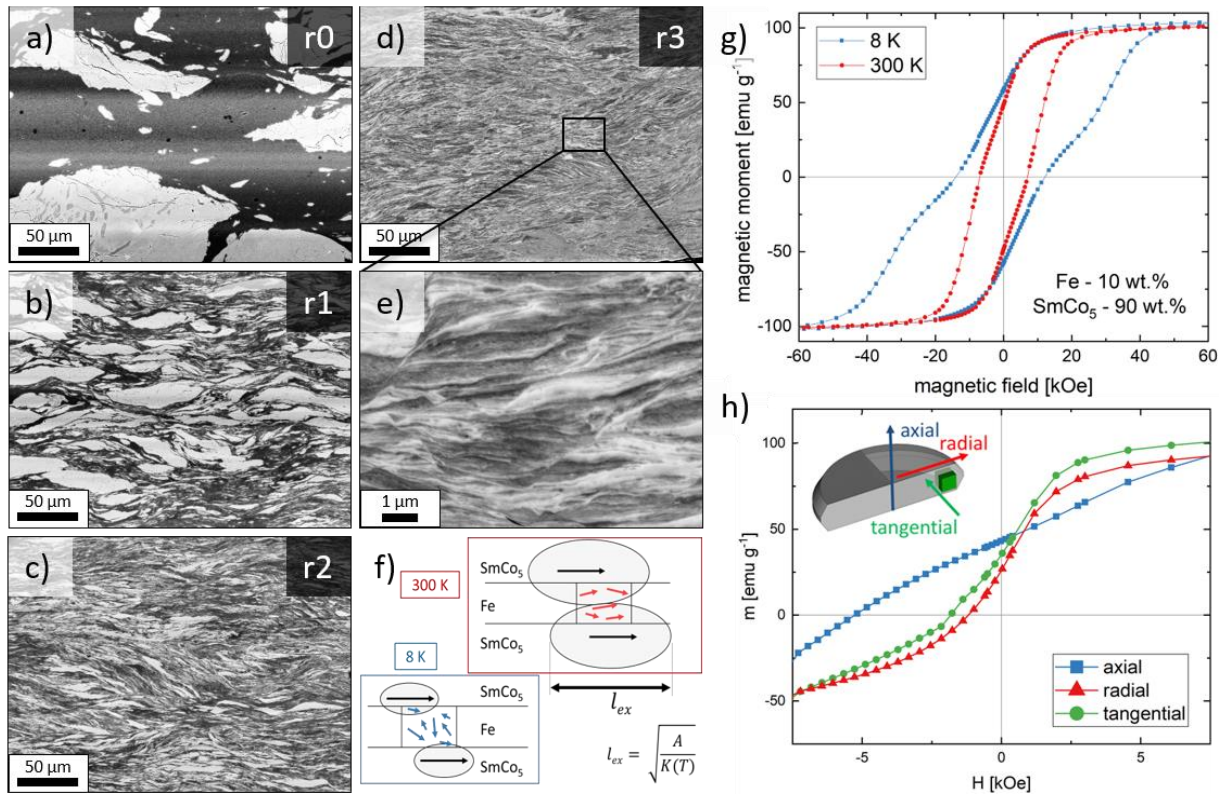

*Figure 9: **a)-d)** BSE-images of an HPT-deformed 47 wt.% Fe - SmCo$_5$ composite at different radii. **e)** shows the lamellar structure at a higher magnification. The temperature dependent exchange length is schematically shown in **f)** with respect to the lamellar morphology. **g)** Changing the temperature, allows to switch between coupling and decoupling of soft and hard magnetic composite. **h)** Magnetic anisotropic behavior induced by the shear strain through HPT.*

As presented in Figure 9g), magnetic hysteresis measurements at different temperatures are performed on a Fe-SmCo$_5$ composite with 10 wt.% Fe and 90 wt.% SmCo$_5$. A different $H_c$, but also a change of the shape of the hysteresis is visible. An exchange coupled behavior at higher temperatures is found and a decoupled compound with magnetically separated phases (pronounced by the 'knee' apparent in recoil curve) is measured at low temperatures. This effect is due to the lamellar microstructure. The temperature dependency of the magnetocrystalline anisotropy of the SmCo$_5$ phase is invers proportional to the exchange length, which is decisively responsible if exchange coupling can occur. Therefore, the Fe phase is not completely coupled to the SmCo$_5$ phase, as the decreasing temperature decreases the exchange length (Figure 9f). Moreover, magnetic anisotropic characteristics (Figure 9h) measured in directions with respect to the HPT-disc are correlated to a crystallographic texture upon deformation.[15] This is further proven by synchrotron diffraction experiments for both the Fe and the SmCo$_5$ phases. The c-axis of the hexagonal SmCo$_5$ orients in axial HPT-disc direction, while the easy axis of the Fe phase points towards the tangential HPT-disc direction. Also Staab et al.[16] reported an deformation induced texture of the SmCo$_5$-Cu nanocomposites. XRD analyses revealed increasing intensities of reflexes containing c-components whereas the intensity of reflexes which did not contain c-components decreased. In addition to XRD analyses, the magnetic measurements conducted along different sample directions, similar to above described, showed this preferential orientation of the c-axis along the axial direction of the HPT disc as well. This reported texture of the SmCo$_5$, crystallizing in a hexagonal structure (CaCu$_5$-type)[45], is of potential interest since it is one of the main challenges in hard magnetic materials processing. The advantage may be the hexagonal crystal structure of SmCo$_5$ which is plastically more anisotropic compared to fcc or bcc materials resulting in a pronounced basal texture during HPT as reported for Mg.[36,129]

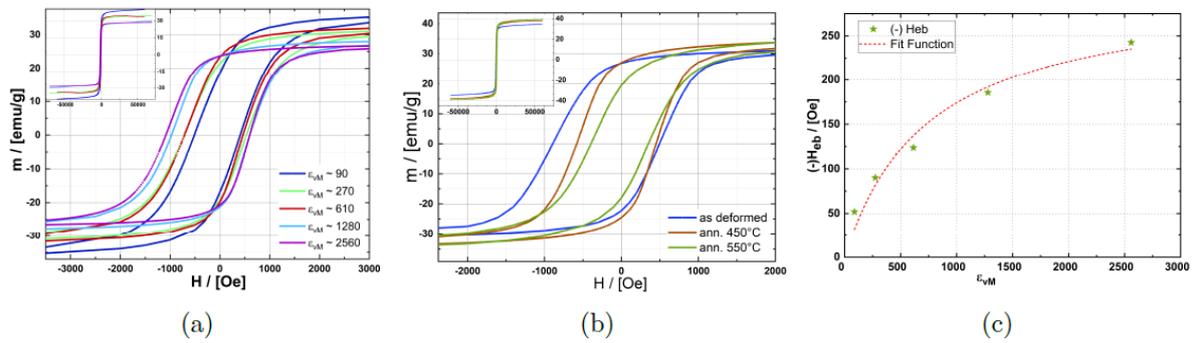

Figure 10: **a)** Hysteresis loops of a Fe$_{10}$Ni$_{40}$-NiO$_{50}$ nanocomposite HPT-deformed to different amounts of applied strain measured at 8 K. **b)** Hysteresis loops of as deformed and annealed Fe$_{10}$Ni$_{40}$-NiO$_{50}$ nanocomposites. **c)** Measured exchange bias for different amounts of applied strain. Reprinted from.[130]

The coercivity and the squareness of the hysteresis loops can be also enhanced at room temperature in nanocomposites consisting of ferromagnetic and antiferromagnetic components with a Neel temperature, $T_N$, above room temperature using the exchange bias effect. The enhancement is maximum after field annealing above $T_N$.[47] The possibility to achieve an exchange bias was recently reported for another type IV material system, the Fe-NiO system. In a first study, the possibility to synthesize nanocomposites by HPT-deformation consisting of Fe (ferromagnetic) and NiO (antiferromagnetic) phases with a phase ratio of 1:1 was investigated. In[130], the synthesis of ferromagnetic-antiferromagnetic nanocomposites by HPT was evaluated to a much greater extent by studying also different ratios of ferromagnetic-antiferromagnetic phases (Fe or Fe$_{20}$Ni$_{80}$ and NiO) and by further investigating the influence on magnetic properties. With rising strain, an increase in exchange bias ($H_{eb}$) was found which could be related to the microstructural refinement of the nanocomposites. As example, the hysteresis loop of an HPT processed Fe$_{10}$Ni$_{40}$-NiO$_{50}$ nanocomposite is shown in Figure 10a). Tuning of magnetic properties was further possible by subsequent annealing treatments (Figure 10b). The fit in Figure 10c) shows a saturating behavior for increasing applied strain for the maximal achievable $H_{eb}$. Menéndez et al.[131] studied a similar type IV system (Co-NiO). While they differentiate between powder and ball milled powder as starting material prior HPT-deformation, the applied strain was relatively low, resulting in a coarser microstructure and lower $H_{eb}$ values.

Phase transformations using HPT was demonstrated by Lee et al.[87] They used and compared arc-melted ingots, powders and nano-sized powders during HPT-deformation, as starting materials to induce the $L1_0$-FeNi phase transformation. Several millions of years are required to naturally form $L1_0$-FeNi. However, enhanced diffusion processes due to HPT-induced lattice defects drastically reduces annealing times to a few days, which is stronger pronounced for fine powders. Using HPT processing for Mn-Al-C based permanent magnets is further proposed by Popov et al.[132] When deforming the τ-phase a strong increase of $H_c$ with increasing radius and thus applied strain is found (from 0.22 T up to 0.58 T). In addition, a stress-driven phase transformation of the metastable τ-phase into the β-phase was reported. They suggested a combination of hot extrusion and HPT to form magnetic anisotropic grains through a textured microstructure and utilize the large structural defect density to foster domain wall pinning.

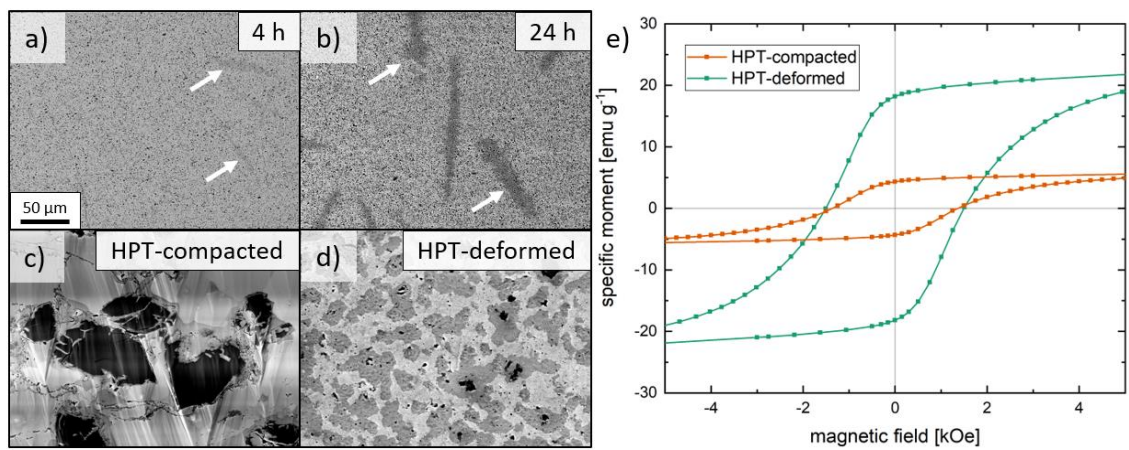

*Figure 11: HPT-deformed MnBi composite after conventional annealing for **a)** 4h and **b)** 24h showing the ferromagnetic α-MnBi phase formation. This phase formation is enhanced after annealing for 4h in presence of an external isostatic magnetic field and for higher deformation grades **c)-d)**. Corresponding magnetic hysteresis curves are shown in **e)**.*

Weissitsch et al.[133] used the HPT process in combination with annealing treatments to circumvent common processing limits within the MnBi material system. The formation of the hard magnetic α-MnBi phase (medium dark contrast, marked by arrows) is reported after equiatomic MnBi powder blends are HPT compacted and annealed for different times as shown in Figure 11a) and b). This intermetallic phase is a potential candidate to substitute rare-earth based magnets, but its formation is a challenging task. They monitored the phase formation by in-situ synchrotron annealing experiments and found an enhanced number of α-MnBi phase nucleation sites for HPT-deformed samples. The α-MnBi phase formation is further improved when a magnetic field is applied during the annealing treatment. This again is strongly dependent on the deformation history. BSE images of an HPT compacted (Figure 11c) and deformed (Figure 11d) sample after the same in-field annealing treatment is shown. The deformed sample, which shows refined Mn particles (dark contrast), also exhibits an extended amount of α-MnBi phase (medium contrast). The saturation magnetization of the composite scales with the existing amount of α-MnBi phase and magnetic SQUID measurements support these results as shown in Figure 11e).

**Summary**

HPT-deformation applied to powder blends of magnetic materials has shown a rising interest in recent years. The process enables several advantages such as grain refinement and the production of nanostructured composite materials or supersaturated solid solutions influencing the magnetic properties. Nevertheless, the process is not limited to obtain technically or scientifically relevant soft- and hard magnetic materials. Reports on superconducting materials show an influence of plastic deformation on the microstructure formation and resulting magnetic properties. Similarities to phase transitions and refined grains are reported.[134–139] Further works on high entropy alloys[140,141] and shape memory materials are found.[142]

The possibilities of the high-pressure torsion process applied to magnetic materials is probably an often underestimated technique. The opportunities of microstructural tuning (i.a., grain refinement, texture evolution or mechanical alloying processes) are combined with sample sizes exhibiting bulk dimensions. In particular, the chance to choose the starting materials which are not limited to bulk but also any kind of powder blends are feasible, allows a variety of possible microstructures and enables sophisticated pathways for material scientists. Besides the starting material itself, process parameters such as pressure, strain and temperature offer a wide range to tune the microstructure and hence the magnetic properties further. As a result, the HPT process provides a great variety of opportunities to further investigate different material systems interesting for magnetic applications.

**Acknowledgement**

This project has received funding from the European Research Council (ERC) under the European Union's Horizon 2020 research and innovation programme (Grant No. 757333 and 101069203).

Funded by the Deutsche Forschungsgemeinschaft (DFG, German Research Foundation) – Project-ID 405553726 – TRR 270, project A08.

**References**

1  K. Edalati, D. Akama, A. Nishio, S. Lee, Y. Yonenaga, J.M. Cubero-Sesin, and Z. Horita: *Acta Materialia*, 2014, vol. 69, pp. 68–77.
2  R. Pippan, S. Scheriau, A. Taylor, M. Hafok, A. Hohenwarter, and A. Bachmaier: *Annu. Rev. Mater. Res.*, 2010, vol. 40, pp. 319–43.
3  K.S. Kormout, R. Pippan, and A. Bachmaier: *Adv. Eng. Mater.*, 2017, vol. 19, p. 1600675.
4  A. Bachmaier and R. Pippan*: *International Materials Reviews*, 2013, vol. 58, pp. 41–62.
5  Y. Ivanisenko, A. Kilmametov, H. Rösner, and R.Z. Valiev: *International Journal of Materials Research*, 2008, vol. 99, pp. 36–41.
6  X. Sauvage, A. Chbihi, and X. Quelennec: *J. Phys.: Conf. Ser.*, 2010, vol. 240, p. 012003.
7  H. Wang, W. Dmowski, Z. Wang, J. Qiang, K. Tsuchiya, Y. Yokoyama, H. Bei, and T. Egami: *Appl. Phys. Lett.*, 2019, vol. 114, p. 061903.
8  B.B. Straumal, A.R. Kilmametov, Y. Ivanisenko, A.A. Mazilkin, O.A. Kogtenkova, L. Kurmanaeva, A. Korneva, P. Zięba, and B. Baretzky: *International Journal of Materials Research*, 2015, vol. 106, pp. 657–64.
9  A. Bachmaier, M. Pfaff, M. Stolpe, H. Aboulfadl, and C. Motz: *Acta Materialia*, 2015, vol. 96, pp. 269–83.
10  D. Gunderov and V. Astanin: *Metals*, 2020, vol. 10, p. 415.
11  K. Edalati and Z. Horita: *Materials Science and Engineering: A*, 2016, vol. 652, pp. 325–52.
12  A. Hohenwarter, A. Bachmaier, B. Gludovatz, S. Scheriau, and R. Pippan: *International Journal of Materials Research*, 2009, vol. 100, pp. 1653–61.
13  J. Cui, J. Ormerod, D. Parker, R. Ott, A. Palasyuk, S. Mccall, M.P. Paranthaman, M.S. Kesler, M.A. McGuire, I.C. Nlebedim, C. Pan, and T. Lograsso: *JOM*, 2022, vol. 74, pp. 1279–95.
14  J. Cui, J. Ormerod, D.S. Parker, R. Ott, A. Palasyuk, S. McCall, M.P. Paranthaman, M.S. Kesler, M.A. McGuire, C. Nlebedim, C. Pan, and T. Lograsso: *JOM*, 2022, vol. 74, pp. 2492–506.
15  L. Weissitsch, M. Stückler, S. Wurster, J. Todt, P. Knoll, H. Krenn, R. Pippan, and A. Bachmaier: *Nanomaterials*, 2022, vol. 12, p. 963.
16  F. Staab, E. Bruder, L. Schäfer, K. Skokov, D. Koch, B. Zingsem, E. Adabifiroozjaei, L. Molina-Luna, O. Gutfleisch, and K. Durst: *Acta Materialia*, 2023, vol. 246, p. 118709.
17  K. Edalati, A. Bachmaier, V.A. Beloshenko, Y. Beygelzimer, V.D. Blank, W.J. Botta, K. Bryła, J. Čížek, S. Divinski, N.A. Enikeev, Y. Estrin, G. Faraji, R.B. Figueiredo, M. Fuji, T. Furuta, T. Grosdidier, J. Gubicza, A. Hohenwarter, Z. Horita, J. Huot, Y. Ikoma, M. Janeček, M. Kawasaki, P. Král, S. Kuramoto, T.G. Langdon, D.R. Leiva, V.I. Levitas, A. Mazilkin, M. Mito, H. Miyamoto, T. Nishizaki, R. Pippan, V.V. Popov, E.N. Popova, G. Purcek, O. Renk, Á. Révész, X. Sauvage, V. Sklenicka, W. Skrotzki, B.B. Straumal, S. Suwas, L.S. Toth, N. Tsuji, R.Z. Valiev, G. Wilde, M.J. Zehetbauer, and X. Zhu: *Materials Research Letters*, 2022, vol. 10, pp. 163–256.
18  R.Z. Valiev, R.K. Islamgaliev, and I.V. Alexandrov: *Progress in Materials Science*, 2000, vol. 45, pp. 103–89.
19  Z.-R. Wang, P.-Z. Si, J. Park, C.-J. Choi, and H.-L. Ge: *Materials*, 2022, vol. 15, p. 2129.
20  A.B. Kustas, D.F. Susan, and T. Monson: *JOM*, 2022, vol. 74, pp. 1306–28.
21  O. Gutfleisch, M.A. Willard, E. Brück, C.H. Chen, S.G. Sankar, and J.P. Liu: *Adv. Mater.*, 2011, vol. 23, pp. 821–42.
22  J.M.D. Coey: *Magnetism and Magnetic Materials*, Cambridge University Press, 2010.
23  J.M.D. Coey: *IEEE Trans. Magn.*, 2011, vol. 47, pp. 4671–81.
24  H. Kronmüller: *phys. stat. sol. (b)*, 1987, vol. 144, pp. 385–96.
25  B. Balasubramanian, P. Mukherjee, R. Skomski, P. Manchanda, B. Das, and D.J. Sellmyer: *Sci Rep*, 2014, vol. 4, p. 6265.
26  R. Goldfarb and F. Fickett: 1985.
27  R. Skomski: *Simple Models of Magnetism*, Oxford University Press, Oxford ; New York, 2008.
28  H.R. Hilzinger: *Appl. Phys.*, 1977, vol. 12, pp. 253–60.

29	F. Bittner, J. Freudenberger, L. Schultz, and T.G. Woodcock: *Journal of Alloys and Compounds*, 2017, vol. 704, pp. 528–36.
30	H. Kronmüller: in *Moderne Probleme der Metallphysik: Zweiter Band Chemische Bindung in Kristallen und Ferromagnetismus*, A. Seeger, ed., Springer, Berlin, Heidelberg, 1966, pp. 24–156.
31	W.F. Brown: *Phys. Rev.*, 1941, vol. 60, pp. 139–47.
32	H. Kronmüller: *Journal of Applied Physics*, 1967, vol. 38, pp. 1314–5.
33	J. Fidler, H. Kirchmayer, and P. Skalicky: *Philosophical Magazine B*, 1981, vol. 43, pp. 765–80.
34	D. Goll and H. Kronmüller: *Naturwissenschaften*, 2000, vol. 87, pp. 423–38.
35	G. Herzer: *Scripta Metallurgica et Materialia*, 1995, vol. 33, pp. 1741–56.
36	H.-J. Lee, S.K. Lee, K.H. Jung, G.A. Lee, B. Ahn, M. Kawasaki, and T.G. Langdon: *Materials Science and Engineering: A*, 2015, vol. 630, pp. 90–8.
37	K. Hono and H. Sepehri-Amin: *Scripta Materialia*, 2012, vol. 67, pp. 530–5.
38	H. Tang, M.A.H. Mamakhel, and M. Christensen: *J. Mater. Chem. C*, 2020, vol. 8, pp. 2109–16.
39	K. Suzuki, N. Ito, S. Saranu, U. Herr, A. Michels, and J.S. Garitaonandia: *Journal of Applied Physics*, 2008, vol. 103, p. 07E730.
40	R. Ramesh and K. Srikrishna: *Journal of Applied Physics*, 1988, vol. 64, pp. 6406–15.
41	R. Ramesh, G. Thomas, and B.M. Ma: *Journal of Applied Physics*, 1988, vol. 64, pp. 6416–23.
42	J. Li, H. Sepehri-Amin, T. Sasaki, T. Ohkubo, and K. Hono: *Science and Technology of Advanced Materials*, 2021, vol. 22, pp. 386–403.
43	K.G. Efthimiadis and N. Ntallis: *Physica B: Condensed Matter*, 2018, vol. 531, pp. 159–63.
44	J. Sung Lee, J. Myung Cha, H. Young Yoon, J.-K. Lee, and Y. Keun Kim: *Sci Rep*, 2015, vol. 5, p. 12135.
45	O. Gutfleisch: *J. Phys. D: Appl. Phys.*, 2000, vol. 33, p. R157.
46	E.F. Kneller and R. Hawig: *IEEE TRANSACTIONS ON MAGNETICS*.
47	J. Nogués, J. Sort, V. Langlais, V. Skumryev, S. Suriñach, J.S. Muñoz, and M.D. Baró: *Physics Reports*, 2005, vol. 422, pp. 65–117.
48	K.P. Skokov and O. Gutfleisch: *Scripta Materialia*, 2018, vol. 154, pp. 289–94.
49	R. Grössinger, R. Sato, D. Holzer, and M. Dahlgren: *Adv. Eng. Mater.*, 2003, vol. 5, pp. 285–90.
50	R.Z. Valiev, R.R. Mulyukov, and V.V. Ovchinnikov: *Philosophical Magazine Letters*, 1990, vol. 62, pp. 253–6.
51	Kh.Ya. Mulyukov, G.F. Korznikova, and R.F. Valiev: *Phys. Stat. Sol. (a)*, 1991, vol. 125, pp. 609–14.
52	S. Scheriau, K. Rumpf, S. Kleber, and R. Pippan: *MSF*, 2008, vol. 584–586, pp. 923–8.
53	S. Scheriau, M. Kriegisch, S. Kleber, N. Mehboob, R. Grössinger, and R. Pippan: *Journal of Magnetism and Magnetic Materials*, 2010, vol. 322, pp. 2984–8.
54	A. Hosokawa, H. Ohtsuka, T. Li, S. Ii, and K. Tsuchiya: *Mater. Trans.*, 2014, vol. 55, pp. 1286–91.
55	Y. Oba, N. Adachi, Y. Todaka, E.P. Gilbert, and H. Mamiya: *Phys. Rev. Research*, 2020, vol. 2, p. 033473.
56	K. Edalati, S. Toh, M. Arita, M. Watanabe, and Z. Horita: *Appl. Phys. Lett.*, 2013, vol. 102, p. 181902.
57	Kh.Ya. Mulyokov, G.F. Korznikova, R.Z. Abdulov, and R.Z. Valiev: *Journal of Magnetism and Magnetic Materials*, 1990, vol. 89, pp. 207–13.
58	R. Grössinger, N. Mehboob, M. Kriegisch, A. Bachmaier, and R. Pippan: *IEEE Trans. Magn.*, 2012, vol. 48, pp. 1473–6.
59	S. Taskaev, K. Skokov, V. Khovaylo, D. Gunderov, and D. Karpenkov: *Physics Procedia*, 2015, vol. 75, pp. 1404–9.
60	A. Chbihi, X. Sauvage, C. Genevois, D. Blavette, D. Gunderov, and A.G. Popov: *Adv. Eng. Mater.*, 2010, vol. 12, pp. 708–13.
61	I. Shchetinin, I. Bordyuzhin, V. Menushenkov, A. Kamynin, A. Savchenko, R. Sundeev, and V. Verbetskii: 2019, pp. 816–20.
62	P.-Z. Si, C.-J. Choi, J. Park, H.-L. Ge, and J. Du: *AIP Advances*, 2020, vol. 10, p. 015320.

63	P.Z. Si, J.T. Lim, J. Park, H.H. Lee, H. Ge, H. Lee, S. Han, H.S. Kim, and C.J. Choi: *Phys. Status Solidi B*, 2020, vol. 257, p. 1900356.
64	B.B. Straumal, A.R. Kilmametov, A.A. Mazilkin, S.G. Protasova, K.I. Kolesnikova, P.B. Straumal, and B. Baretzky: *Materials Letters*, 2015, vol. 145, pp. 63–6.
65	V.V. Stolyarov, D.V. Gunderov, A.G. Popov, V.S. Gaviko, and A.S. Ermolenko: *Journal of Alloys and Compounds*, 1998, vol. 281, pp. 69–71.
66	I.S. Tereshina, I.A. Pelevin, E.A. Tereshina, G.S. Burkhanov, K. Rogacki, M. Miller, N.V. Kudrevatykh, P.E. Markin, A.S. Volegov, R.M. Grechishkin, S.V. Dobatkin, and L. Schultz: *Journal of Alloys and Compounds*, 2016, vol. 681, pp. 555–60.
67	B.B. Straumal, A.A. Mazilkin, S.G. Protasova, D.V. Gunderov, G.A. López, and B. Baretzky: *Materials Letters*, 2015, vol. 161, pp. 735–9.
68	A.G. Popov, V.S. Gaviko, N.N. Shchegoleva, L.A. Shreder, D.V. Gunderov, V.V. Stolyarov, W. Li, L.L. Li, and X.Y. Zhang: *Journal of Iron and Steel Research, International*, 2006, vol. 13, pp. 160–5.
69	D.V. Gunderov and V.V. Stolyarov: *Journal of Applied Physics*, 2010, vol. 108, p. 053901.
70	W. Li, L. Li, Y. Nan, Z. Xu, X. Zhang, A.G. Popov, D.V. Gunderov, and V.V. Stolyarov: *Journal of Applied Physics*, 2008, vol. 104, p. 023912.
71	W. Li, L. Li, Y. Nan, X. Li, X. Zhang, D.V. Gunderov, V.V. Stolyarov, and A.G. Popov: *Appl. Phys. Lett.*, 2007, vol. 91, p. 062509.
72	H. Li, W. Li, Y. Zhang, D.V. Gunderov, and X. Zhang: *Journal of Alloys and Compounds*, 2015, vol. 651, pp. 434–9.
73	A. Musiał, Z. Śniadecki, N. Pierunek, Yu. Ivanisenko, D. Wang, M.H. Fawey, and B. Idzikowski: *Journal of Alloys and Compounds*, 2019, vol. 787, pp. 794–800.
74	X. Li, L. Lou, W. Song, Q. Zhang, G. Huang, Y. Hua, H.-T. Zhang, J. Xiao, B. Wen, and X. Zhang: *Nano Lett.*, 2017, vol. 17, pp. 2985–93.
75	H. Li, L. Lou, F. Hou, D. Guo, W. Li, X. Li, D.V. Gunderov, K. Sato, and X. Zhang: *Appl. Phys. Lett.*, 2013, vol. 103, p. 142406.
76	H. Li, W. Li, D. Guo, and X. Zhang: *Journal of Magnetism and Magnetic Materials*, 2017, vol. 425, pp. 84–9.
77	A.G. Popov, V.S. Gaviko, N.N. Shchegoleva, L.A. Shreder, V.V. Stolyarov, D.V. Gunderov, X.Y. Zhang, W. Li, and L.L. Li: *Phys. Metals Metallogr.*, 2007, vol. 104, pp. 238–47.
78	A.G. Popov, V.V. Serikov, and N.M. Kleinerman: *Phys. Metals Metallogr.*, 2010, vol. 109, pp. 505–13.
79	I.V. Shchetinin, I.G. Bordyuzhin, R.V. Sundeev, V.P. Menushenkov, A.V. Kamynin, V.N. Verbetsky, and A.G. Savchenko: *Materials Letters*, 2020, vol. 274, p. 127993.
80	A. Hosokawa and Y. Hirayama: *Scripta Materialia*, 2023, vol. 226, p. 115251.
81	M. Ulyanov, S. Taskaev, S. Shevyrtalov, P. Medvedskaya, and D. Gunderov: *AIP Advances*, 2021, vol. 11, p. 025311.
82	M.N. Ulyanov, M.Yu. Bogush, M.A. Gavrilova, S.V. Taskaev, Z. Hu, D.V. Gunderov, D.A. Zherebtsov, and E.P. Ulyanova: *Челябинский физико-математический журнал*, 2021, vol. 6, pp. 255–63.
83	S.V. Taskaev, M.N. Ulyanov, D.V. Gunderov, and M.Yu. Bogush: *IEEE Magn. Lett.*, 2020, vol. 11, pp. 1–4.
84	M. Kołodziej, Z. Śniadecki, A. Musiał, N. Pierunek, Yu. Ivanisenko, A. Muszyński, and B. Idzikowski: *Journal of Magnetism and Magnetic Materials*, 2020, vol. 502, p. 166577.
85	S.W. Lee and Z. Horita: *MSF*, 2010, vol. 667–669, pp. 313–8.
86	T. Ohtsuki, M. Kotsugi, T. Ohkochi, S. Lee, Z. Horita, and K. Takanashi: *Journal of Applied Physics*, 2013, vol. 114, p. 143905.
87	S. Lee, K. Edalati, H. Iwaoka, Z. Horita, T. Ohtsuki, T. Ohkochi, M. Kotsugi, T. Kojima, M. Mizuguchi, and K. Takanashi: *Philosophical Magazine Letters*, 2014, vol. 94, pp. 639–46.
88	M.V. Gorshenkov, D.Y. Karpenkov, R.V. Sundeev, V.V. Cheverikin, and I.V. Shchetinin: *Materials Letters*, 2020, vol. 272, p. 127864.
89	A.S. Fortuna, M.V. Gorshenkov, and R.V. Sundeev: *Materials Letters*, 2021, vol. 296, p. 129888.

90   A.S. Fortuna, M.V. Gorshenkov, V.V. Cheverikin, and R.V. Sundeev: *Journal of Alloys and Compounds*, 2022, vol. 901, p. 163424.
91   Y. Jia, Y. Wu, S. Zhao, S. Zuo, K.P. Skokov, O. Gutfleisch, C. Jiang, and H. Xu: *Phys. Rev. Materials*, 2020, vol. 4, p. 094402.
92   Y. Jia, Y. Wu, Y. Xu, R. Zheng, S. Zhao, K.P. Skokov, F. Maccari, A. Aubert, O. Gutfleisch, J. Wang, H. Wang, J. Zou, and C. Jiang: *Acta Materialia*, 2023, vol. 245, p. 118654.
93   P.W. Bridgman: *Phys. Rev.*, 1935, vol. 48, pp. 825–47.
94   C.M. Boos: *GEOPHYSICS*, 1936, vol. 1, pp. 379–80.
95   E.Y. Yoon, D.J. Lee, D.-H. Ahn, E.S. Lee, and H.S. Kim: *J Mater Sci*, 2012, vol. 47, pp. 7770–6.
96   A.V. Korznikov, I.M. Safarov, D.V. Laptionok, and R.Z. Valiev: *Acta Metallurgica et Materialia*, 1991, vol. 39, pp. 3193–7.
97   A. Bachmaier, A. Hohenwarter, and R. Pippan: *Scripta Materialia*, 2009, vol. 61, pp. 1016–9.
98   I.V. Alexandrov, R.K. Islamgaliev, R.Z. Valiev, Y.T. Zhu, and T.C. Lowe: *Metall Mater Trans A*, 1998, vol. 29, pp. 2253–60.
99   H. Li, A. Misra, Z. Horita, C.C. Koch, N.A. Mara, P.O. Dickerson, and Y. Zhu: *Appl. Phys. Lett.*, 2009, vol. 95, p. 071907.
100  M. Wurmshuber, S. Doppermann, S. Wurster, and D. Kiener: *IOP Conf. Ser.: Mater. Sci. Eng.*, 2019, vol. 580, p. 012051.
101  M. Stückler, J. Zálešák, T. Müller, S. Wurster, L. Weissitsch, M. Meier, P. Felfer, C. Gammer, R. Pippan, and A. Bachmaier: *Journal of Alloys and Compounds*, 2022, vol. 901, p. 163616.
102  L. Weissitsch, M. Stückler, S. Wurster, R. Pippan, and A. Bachmaier: *MSF*, 2021, vol. 1016, pp. 1603–10.
103  K. Edalati: *Advanced Engineering Materials*, 2019, vol. 21, p. 1800272.
104  M. Stückler, H. Krenn, R. Pippan, L. Weissitsch, S. Wurster, and A. Bachmaier: *Nanomaterials*, 2018, vol. 9, p. 6.
105  A. Bachmaier, H. Krenn, P. Knoll, H. Aboulfadl, and R. Pippan: *Journal of Alloys and Compounds*, 2017, vol. 725, pp. 744–9.
106  M. Stückler, H. Krenn, P. Kürnsteiner, B. Gault, F. De Geuser, L. Weissitsch, S. Wurster, R. Pippan, and A. Bachmaier: *Acta Materialia*, 2020, vol. 196, pp. 210–9.
107  M. Stückler, L. Weissitsch, S. Wurster, P. Felfer, H. Krenn, R. Pippan, and A. Bachmaier: *AIP Advances*, 2020, vol. 10, p. 015210.
108  J.R. Childress and C.L. Chien: *Phys. Rev. B*, 1991, vol. 43, pp. 8089–93.
109  H. Kronmüller and M. Fähnle: *Micromagnetism and the Microstructure of Ferromagnetic Solids*, 2003.
110  T.D. Shen, R.B. Schwarz, and J.D. Thompson: *Phys. Rev. B*, 2005, vol. 72, p. 014431.
111  R. Moreno, R.F.L. Evans, S. Khmelevskyi, M.C. Muñoz, R.W. Chantrell, and O. Chubykalo-Fesenko: *Phys. Rev. B*, 2016, vol. 94, p. 104433.
112  C.L. Chien, S.H. Liou, D. Kofalt, W. Yu, T. Egami, T.J. Watson, and T.R. McGuire: *Phys. Rev. B*, 1986, vol. 33, pp. 3247–50.
113  P. Crespo, I. Navarro, A. Hernando, P. Rodríguez, A.G. Escorial, J.M. Barandiarán, O. Drbohlav, and A.R. Yavari: *Journal of Magnetism and Magnetic Materials*, 1995, vol. 150, pp. 409–16.
114  T. Ambrose, A. Gavrin, and C.L. Chien: *Journal of Magnetism and Magnetic Materials*, 1993, vol. 124, pp. 15–9.
115  O. Schneeweiss, M. Friák, M. Dudová, D. Holec, M. Šob, D. Kriegner, V. Holý, P. Beran, E.P. George, J. Neugebauer, and A. Dlouhý: *Phys. Rev. B*, 2017, vol. 96, p. 014437.
116  C.L. Chien, S.H. Liou, G. Xiao, and M.A. Gatzke: *MRS Proc.*, 1986, vol. 80, p. 395.
117  M. Stückler, L. Weissitsch, S. Wurster, H. Krenn, R. Pippan, and A. Bachmaier: *Journal of Materials Research and Technology*, 2021, vol. 12, pp. 1235–42.
118  M. Stückler, S. Wurster, R. Pippan, and A. Bachmaier: *AIP Advances*, 2021, vol. 11, p. 015033.
119  G. Binasch, P. Grünberg, F. Saurenbach, and W. Zinn: *Phys. Rev. B*, 1989, vol. 39, pp. 4828–30.
120  M.N. Baibich, J.M. Broto, A. Fert, F.N. Van Dau, F. Petroff, P. Etienne, G. Creuzet, A. Friederich, and J. Chazelas: *Phys. Rev. Lett.*, 1988, vol. 61, pp. 2472–5.
121  J.Q. Xiao, J.S. Jiang, and C.L. Chien: *Phys. Rev. Lett.*, 1992, vol. 68, pp. 3749–52.

122  A.E. Berkowitz, J.R. Mitchell, M.J. Carey, A.P. Young, S. Zhang, F.E. Spada, F.T. Parker, A. Hutten, and G. Thomas: *Phys. Rev. Lett.*, 1992, vol. 68, pp. 3745–8.
123  S. Wurster, L. Weissitsch, M. Stückler, P. Knoll, H. Krenn, R. Pippan, and A. Bachmaier: *Metals*, 2019, vol. 9, p. 1188.
124  S. Wurster, M. Stückler, L. Weissitsch, T. Müller, and A. Bachmaier: *Applied Sciences*, 2020, vol. 10, p. 5094.
125  M. Kasalo, S. Wurster, M. Stückler, M. Zawodzki, L. Weissitsch, R. Pippan, and A. Bachmaier: *IOP Conf. Ser.: Mater. Sci. Eng.*, 2022, vol. 1249, p. 012047.
126  D.V. Gunderov, A.G. Popov, N.N. Schegoleva, V.V. Stolyarov, and A.R. Yavary: in *Nanomaterials by Severe Plastic Deformation*, M. Zehetbauer and R.Z. Valiev, eds., Wiley-VCH Verlag GmbH & Co. KGaA, Weinheim, FRG, 2005, pp. 165–9.
127  F. Hou, Y. Hua, G. Zhang, M. Li, L. Lou, Q. Zhang, G. Huang, W. Li, and X. Li: *Journal of Magnetism and Magnetic Materials*, 2020, vol. 499, p. 166271.
128  L. Weissitsch, M. Stückler, S. Wurster, P. Knoll, H. Krenn, R. Pippan, and A. Bachmaier: *Crystals*, 2020, vol. 10, p. 1026.
129  B.J. Bonarski, E. Schafler, B. Mikułowski, and M. Zehetbauer: *MSF*, 2008, vol. 584–586, pp. 263–8.
130  M. Zawodzki, L. Weissitsch, H. Krenn, S. Wurster, and A. Bachmaier: *Nanomaterials*, 2023, vol. 13, p. 344.
131  E. Menéndez, J. Sort, V. Langlais, A. Zhilyaev, J.S. Muñoz, S. Suriñach, J. Nogués, and M.D. Baró: *Journal of Alloys and Compounds*, 2007, vol. 434–435, pp. 505–8.
132  V.V. Popov, F. Maccari, I.A. Radulov, A. Kovalevsky, A. Katz-Demyanetz, and M. Bamberger: *Manufacturing Rev.*, 2021, vol. 8, p. 10.
133  L. Weissitsch, S. Wurster, M. Stückler, T. Müller, H. Krenn, R. Pippan, and A. Bachmaier: *submitted to Acta Materialia*.
134  V.A. Beloshenko and V.V. Chishko: *Phys. Metals Metallogr.*, 2013, vol. 114, pp. 992–1002.
135  K. Edalati, T. Daio, S. Lee, Z. Horita, T. Nishizaki, T. Akune, T. Nojima, and T. Sasaki: *Acta Materialia*, 2014, vol. 80, pp. 149–58.
136  M. Mito, H. Matsui, K. Tsuruta, T. Yamaguchi, K. Nakamura, H. Deguchi, N. Shirakawa, H. Adachi, T. Yamasaki, H. Iwaoka, Y. Ikoma, and Z. Horita: *Sci Rep*, 2016, vol. 6, p. 36337.
137  T. Nishizaki, S. Lee, Z. Horita, T. Sasaki, and N. Kobayashi: *Physica C: Superconductivity*, 2013, vol. 493, pp. 132–5.
138  M.J.R. Sandim, D. Stamopoulos, H.R.Z. Sandim, L. Ghivelder, L. Thilly, V. Vidal, F. Lecouturier, and D. Raabe: *Supercond. Sci. Technol.*, 2006, vol. 19, pp. 1233–9.
139  M. Ullrich, A. Leenders, J. Krelaus, L.-O. Kautschor, H.C. Freyhardt, L. Schmidt, F. Sandiumenge, and X. Obradors: *Materials Science and Engineering: B*, 1998, vol. 53, pp. 143–8.
140  W. Wu, M. Song, S. Ni, J. Wang, Y. Liu, B. Liu, and X. Liao: *Sci Rep*, 2017, vol. 7, p. 46720.
141  N. Shkodich, F. Staab, M. Spasova, K.V. Kuskov, K. Durst, and M. Farle: *Materials*, 2022, vol. 15, p. 7214.
142  R. Chulist, W. Skrotzki, C.-G. Oertel, A. Böhm, T. Lippmann, and E. Rybacki: *Scripta Materialia*, 2010, vol. 62, pp. 650–3.